# Structure, Stability and Mechanical Properties of Boron-Rich Mo–B Phases: A Computational Study

*Dmitry V. Rybkovskiy,*[1,2] *Alexander G. Kvashnin,*[1,3,*] *Yulia A. Kvashnina,*[4] *and Artem R. Oganov*[1,3,5]

[1] Skolkovo Institute of Science and Technology, Skolkovo Innovation Center, 3 Nobel Street, Moscow 121205, Russia

[2] A. M. Prokhorov General Physics Institute of RAS, 38 Vavilov Street, Moscow 119991, Russia

[3] Moscow Institute of Physics and Technology, 9 Institutsky Lane, Dolgoprudny 141700, Russia

[4] Pirogov Russian National Research Medical University, 1 Ostrovityanova Street, Moscow 117997, Russia

[5] International Center for Materials Discovery, Northwestern Polytechnical University, Xi'an 710072, China

ABSTRACT: Molybdenum borides were studied theoretically using first-principles calculations, empirical total energy model and global optimization techniques to determine stable crystal structures. Our calculations reveal the structures of known Mo–B phases, attaining close agreement with experiment. Following our developed lattice model, we describe in detail the crystal structure of boron-rich $MoB_x$ phases with $3 \leq x \leq 9$ as the hexagonal $P6_3/mmc$-$MoB_3$ structure with Mo atoms partially replaced by triangular boron units. The most energetically stable arrangement of these $B_3$ units corresponds to their uniform distribution in the bulk of the crystal structure, which leads to the formation of a disordered nonstoichiometric phase, with ordering arising at compositions close to $x=5$ due to a strong repulsive interaction between neighboring $B_3$ units. The most energetically favorable structures of $MoB_x$ correspond to the compositions $4 \lesssim x \leq 5$, with $MoB_5$ being the boron-richest stable phase. The estimated hardness of $MoB_5$ is 37-39 GPa, suggesting that the boron-rich phases are potentially superhard.

Transition metal borides are often considered possible replacements of the traditional superhard materials in various technological applications. Unlike widely used diamond and cubic boron nitride, transition metal borides do not require high pressure for their synthesis, making their production cheaper and more readily scalable. The outstanding mechanical properties of these materials originate from the combination of high valence electron density of metal atoms, responsible for high incompressibility, and covalent bonds of light boron atoms, resisting elastic and plastic shape deformations [1,2]. The borides of tungsten and molybdenum are of particular interest because of their capability to accumulate a significant amount of boron, resulting in high hardness. The W–B system has been extensively studied both experimentally [3–6] and theoretically [7–9], whereas the exact crystal structure of higher Mo–B phases is still under debate.

Today, the experimental methods of structure determination are strongly supported by state-of-the-art computational techniques based on global optimization algorithms and robust quantum-chemical total energy computation methods [10]. Such approaches have already been applied to the investigation of the Mo–B phase diagram, shedding light on some aspects of the crystal structure of molybdenum borides. Liang et al. [11] computed the energies of the experimentally claimed and hypothetical crystal structures. Zhang et al. [12] used the evolutionary global optimization algorithm USPEX for an unbiased study of Mo–B phases. The latter work was, however, restricted to specific chemical compositions. Despite the success of the existing theoretical studies, the structure of the molybdenum borides with compositions close to $MoB_4$, which are of the most interest for potential applications as hard materials, is still controversial. The experimentally proposed structures are mainly based on X-ray diffraction (XRD) experiments, where the largest part of the signal is determined by the positions of Mo atoms, and the correct placement of much lighter boron atoms is complicated. Total energy computations show that such structures have high energies of formation, indicating their instability [11,12]. On the other hand, the theoretically predicted *P*6$_3$/*mmc*-MoB$_4$ structure [12], although having a low energy of formation, has a X-Ray diffraction (XRD) pattern, which is incompatible with what is suggested by experiment. These shortcomings motivate a further search for boron-rich Mo–B phases that would match the experimental XRD and exhibit an energetically preferable boron arrangement.

In this work we use a broad range of computational techniques, including the evolutionary crystal structure prediction to search for thermodynamically stable phases through all possible compositions, density-functional total energy computations, and a parameterized lattice model for the theoretical study of the boron-rich part of the Mo–B phase diagram. We show that within the compositional range between MoB$_4$ and MoB$_5$, the molybdenum borides may be described by a structural model similar to the one proposed by Lech et al. [6] for the highest tungsten boride. On the basis of total energy computations and relative stabilities, we describe the requirements that the structures have to satisfy to be energetically preferable.

Global optimization of Mo–B crystal structures was performed using the variable-composition evolutionary algorithm as implemented in the USPEX code [13–15]. This approach makes it possible to perform an automatic structure search within the whole compositional range of a multicomponent system. During the structure search, the first generation of 120 structures was produced randomly with up to 24 atoms (for the variable-composition search) and 36 atoms (for the fixed-composition search) in the primitive cell. The succeeding generations were obtained by applying the heredity (40% of each generation), softmutation (20%), and transmutation (20%) operators [13–15]; 20% of each generation were produced using random [16] and random topological generators [17].

For the structure relaxations and total energy computations, the projector augmented-wave [18,19] density functional theory (PAW-DFT) was used as implemented within the VASP [20–22] package. The generalized gradient approximation of Perdew–Burke–Ernzerhof (GGA-PBE) [23] was used for the exchange-correlation functional. PAW datasets were used to describe the electron-ion interactions with Mo 4$p$, 4$d$, 5$s$ and B 2$s$, 2$p$ electrons treated as valence electrons. The plane-wave energy cutoff of 400 eV, the Methfessel–Paxton smearing [24] of electronic occupations, and Γ-centered $k$-point meshes with a resolution of $2\pi \times 0.025$ Å$^{-1}$ for the Brillouin zone sampling ensured the convergence of total energies.

To investigate the boron-rich part of the Mo–B phase diagram, a lattice model for the total energy was developed (see Results and Discussion section). X-ray diffraction has been simulated with the VESTA software [25] using 1.54059 Å wavelength.

Vickers hardness was estimated according to Mazhnik-Oganov [26] ($H_V^{MO}$) and Chen's [27] ($H_V^C$) models. Test calculations of Vickers hardness for a number of materials using Mazhnik-Oganov and Chen's models agree well with the reference experimental data: diamond – 99 and 94 GPa (~96 GPa [28]), TiN – 21 and 23 GPa (20.5 GPa [29]), c-BN – 71 and 63 GPa (~66 GPa [30]).

Fracture toughness was calculated using empirical Mazhnik-Oganov model [26] ($K_{IC}^{MO}$) and Niu-Oganov model [31] ($K_{IC}$). Calculated values of fracture toughness for well-studied materials nicely agree with experimental data. Niu-Oganov model gives fracture toughness of diamond, WC, TiN and c-BN equal to 6.3, 5.4, 3.3 and 5.4 MPa·m$^{0.5}$, respectively. Mazhnik-Oganov model gives 6.2, 7.7, 3.8 and 5.4 MPa·m$^{0.5}$ for diamond, WC, TiN and c-BN, respectively. The experimental data are 4-7 MPa·m$^{0.5}$ [32–34] for diamond, 5-8 MPa·m$^{0.5}$ [35,36] for WC, 3.5-5 MPa·m$^{0.5}$ [35] for TiN and 2-5 MPa·m$^{0.5}$ [33,37] for *c*-BN.

**Variable-Composition Evolutionary Structure Search**
To search for stable Mo–B compounds, we used the variable-composition evolutionary algorithm USPEX. A thermodynamically stable phase has a lower energy of formation than any phase or phase assemblage of the same composition and is located on the convex hull line. The calculated energies of formation Δ$E_{form}$, of the stable and metastable structures, obtained during evolutionary search and proposed earlier for some Mo–B phases in Refs. [3,38–41], are shown in Fig. 1 as a function of the atomic fraction of boron.

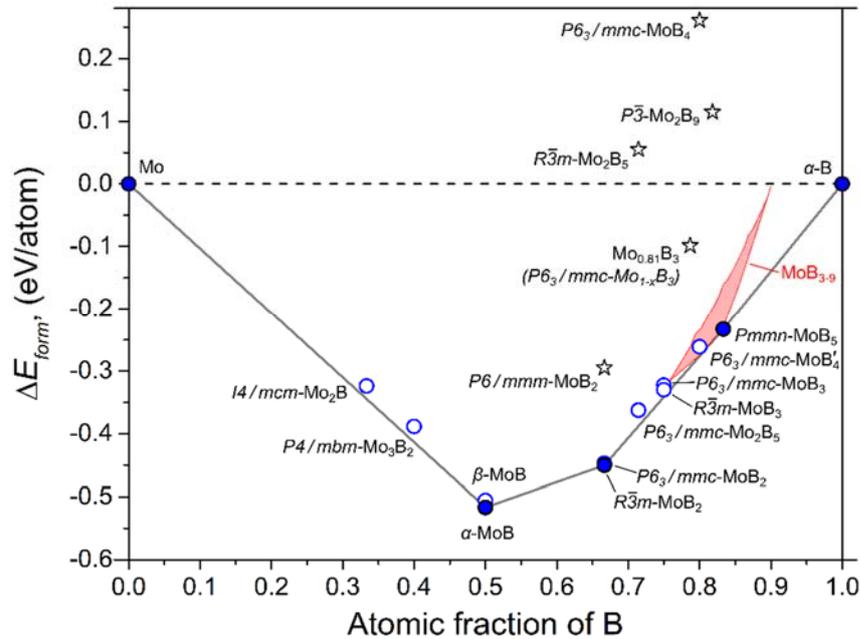

Fig. 1. Calculated energies of formation for the Mo–B system. Filled and hollow circles designate stable and metastable phases, respectively. Stars show the energies of formation of the Mo–B structures proposed in the experimental works ($P6/mmm$-MoB$_2$ [38,42], $R\bar{3}m$-Mo$_2$B$_5$ [3], $P6_3/mmc$-Mo$_{0.81}$B$_3$ of the Mo$_{1-x}$B$_3$ type [39], $P6_3/mmc$-MoB$_4$ [40], and $R\bar{3}m$-Mo$_2$B$_9$ of the W$_{2-x}$B$_9$ type [41]). The shaded area depicts the energies of formation of boron-rich MoB$_x$ phases with $3 \leq x \leq 9$, obtained using a parameterized lattice model.

The only stable phases obtained using the evolutionary search were $I4_1/amd$-MoB (α-MoB), $R\bar{3}m$-MoB$_2$, and $Pmmn$-MoB$_5$. It should be mentioned, however, that vibrational effects might introduce changes into the convex hull diagram at high temperatures, used during synthesis of the molybdenum borides. Such vibrational contributions lead to free energy change of all phases, resulting in the rearrangement of the energies of formation and may result in stabilization of some phases close to the convex hull line. It is therefore important to consider not only systems, lying on the convex hull in Fig. 1, but also metastable phases, which are close to it. Such phases include $I4/mcm$-Mo$_2$B, $P4/mbm$-Mo$_3$B$_2$, $Cmcm$-MoB (β-MoB), $P6_3/mmc$-MoB$_2$, $P6/mmm$-MoB$_2$, $P6_3/mmc$-Mo$_2$B$_5$, $R\bar{3}m$-MoB$_3$, and $P6_3/mmc$-MoB$'_4$, and further discussion on the basis of available experimental and theoretical data from literature will show, that some of them may indeed become stable at finite temperature. The crystal structures of the USPEX-obtained Mo–B phases are shown in Fig. 2. Each of the phases is discussed separately, from those having a low boron content to the boron-richest ones.

The calculated $Mo_2B$ (a $CuAl_2$ prototype) belongs to *I4/mcm* space group and is well-known from the experimental studies [3,38,42,43]. The calculated energy of formation of this phase lies 20 meV/atom above the convex hull, indicating that it is metastable at 0 K, which agrees with the previous theoretical findings [11,12]. We can therefore conclude that vibrational effects are responsible for the stabilization of this phase. Another metastable structure, which has been experimentally found at high temperatures ($T > 1550$ K), is *P4/mbm*-$Mo_3B_2$, whose energy of formation lies 25 meV/atom above the convex hull. It was first reported by Steinitz et al. in 1952 [38], though some later works doubted its existence [42].

We also obtained two known MoB phases: *I4$_1$/amd*-MoB (α-MoB) and *Cmcm*-MoB (β-MoB). The energy of formation of β-MoB is higher than that of α-MoB by 11 meV/atom. *I4$_1$/amd*-MoB undergoes a phase transition to the high-temperature *Cmcm*-MoB at ~1900 K, as known from experiment and predicted by theory [11,38,42].

The chemical compositions and crystal structures of boron-rich Mo–B phases were debated for a long time. We obtained two low-energy structures for $MoB_2$: $R\bar{3}m$-$MoB_2$ and *P6$_3$/mmc*-$MoB_2$ (Fig. 1). The synthesis of $MoB_2$ having the $AlB_2$-type structure (*P6/mmm* space group) [42] was first reported by Steinitz et al. [38]. Later studies by Higashi et al. [44] reported the $R\bar{3}m$ phase. Our calculations show that the energy of formation of *P6/mmm*-$MoB_2$ is higher by 152 meV/atom compared to that of $R\bar{3}m$-$MoB_2$, which agrees with the previous theoretical studies [11,12]. The structure of *P6$_3$/mmc*-$MoB_2$ corresponds to the $hP12$-$WB_2$ type proposed in Ref. [11] and lies 3 meV/atom above the lowest-energy $R\bar{3}m$-$MoB_2$.

$Mo_2B_5$ was reported in several experimental works [3,38,42] and was first described by Kiessling [3] as an $R\bar{3}m$ structure with alternating planar hexagonal and puckered (with additional B atoms) hexagonal boron layers separated by Mo layers [3,43]. The existence of this phase has been questioned by the experiments [44,45], and a conclusion has been made that it is a misinterpreted *hR*18-Mo$_2$B$_4$ structure with $R\bar{3}m$ space group. Later, another metastable, lower-energy *P*6$_3$/*mmc*-Mo$_2$B$_5$ with a different structure was predicted theoretically [12]. We found that $R\bar{3}m$-Mo$_2$B$_5$ is unstable, having a positive energy of formation of 55 meV/atom, and *P*6$_3$/*mmc*-Mo$_2$B$_5$ is 25 meV/atom above the convex hull, which agrees with the previous theoretical data [11,12].

We found that $R\bar{3}m$-MoB$_3$ is metastable, lying 12 meV/atom above the convex hull, which agrees with the previous theoretical results. In comparison, *P*6$_3$/*mmc*-MoB$_3$ differs by the mutual position of the hexagonal Mo layers, and its calculated energy is 7 meV/atom higher, but previous theoretical investigations have suggested that it becomes more stable than the $R\bar{3}m$ phase at high temperatures [11].

The MoB$_4$ structure with *P*6$_3$/*mmc* space group, which we denote as the *P*6$_3$/*mmc*-MoB$'_4$ phase, is metastable, lying 15 meV/atom above the convex hull. This phase is equivalent to those obtained in previous theoretical global optimization studies [12]. It is composed of Mo layers sandwiched between two-atom-thick puckered boron layers. Numerous experimental works reported the synthesis of boron-rich compounds with the compositions close to MoB$_4$ [39,40,46]. However, the structures proposed in these works have a different XRD pattern in comparison to the theoretical *P*6$_3$/*mmc*-MoB$'_4$ phase.

Our evolutionary structure search revealed a new stable boron-rich phase *Pmmn*-MoB$_5$ with the same crystal structure type as that of an earlier reported WB$_5$ [8]. This phase is made of edge- and face-sharing MoB$_{12}$ hexagonal prisms and open B$_{15}$ clusters linked into a 3D structure by B–B bonds. The calculated phonon spectrum shows the dynamical stability of this phase (Supporting Information Fig. S3). The crystal structures of all low-energy phases are presented in Supporting Information Table S1.

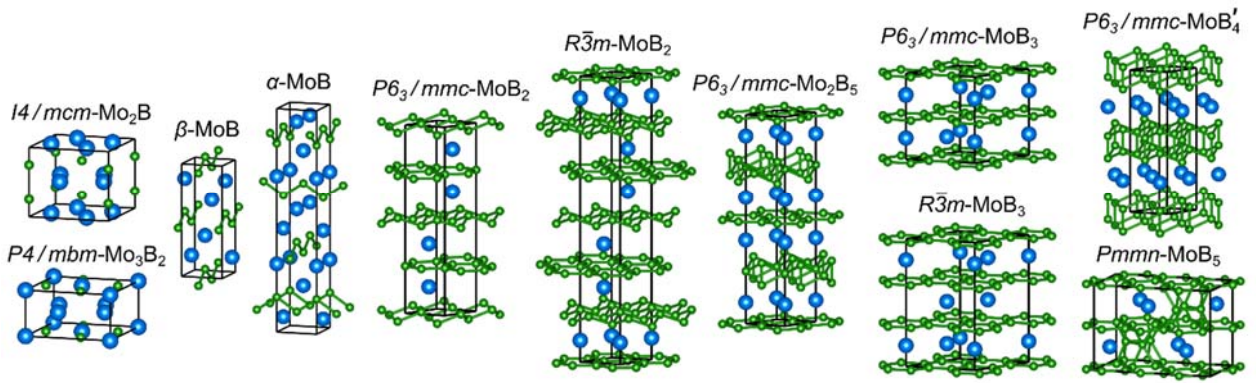

Fig. 2. Crystal structures of the Mo–B phases obtained using the evolutionary crystal structure search.

Because the main practical interest of boron-rich molybdenum borides is the manufacture of hard materials for technological applications, we estimated the mechanical properties of the obtained low-energy Mo–B phases. The results are summarized in Table 1. The Vickers hardness, estimated using the Chen's ($H_V^C$) [27] and Mazhnik-Oganov ($H_V^{MO}$) [26] models show good agreement with the experimental values for α-MoB and β-MoB [38], whereas some overestimation is observed in the case of the $R\bar{3}m$-MoB2 phase [47]. The estimated hardness values for MoB3 are ~33-36 GPa. For the boron richest phase Pmmn-MoB5 we estimated the Vickers hardness as ~37-39 GPa, which is close to the superhard materials boundary of 40 GPa [29,48,49]. Therefore, the highest borides of molybdenum are potentially superhard. It is worth noting, that a recent experimental study of boron-rich molybdenum boride with composition Mo0.757B3 reported anisotropic nanoindentation hardness of ~34-40 GPa [50]. The comparison of mechanical characteristics with known materials is shown in Figure S1 (see Supporting Information).

Table 1. Mechanical properties of the low-energy Mo–B phases

| Phase | B, GPa | G, GPa | E, GPa | $v$ | $H_V^C$, GPa | $H_V^{MO}$, GPa | $H_V^{exp}$, GPa | $K_{IC}$, MPa×m$^{0.5}$ | $K_{IC}^{MO}$, MPa×m$^{0.5}$ |
|---|---|---|---|---|---|---|---|---|---|
| $I4_1/amd$-MoB | 308 | 203 | 500 | 0.229 | 24.5 | 23.2 | 23.0 [38] | 3.9 | 4.2 |
| $Cmcm$-MoB | 307 | 193 | 478 | 0.241 | 22.1 | 21.9 | 24.5 [38] | 3.2 | 4.1 |
| $R\bar{3}m$-MoB2 | 299 | 230 | 548 | 0.193 | 32.4 | 30.3 | 24.2 [47] | 4.7 | 4.2 |

| Phase | B, GPa | G, GPa | E, GPa | $v$ | $H_V^C$, GPa | $H_V^{MO}$, GPa | $H_V^{exp}$, GPa | $K_{IC}$, MPa×m$^{0.5}$ | $K_{IC}^{MO}$, MPa×m$^{0.5}$ |
|---|---|---|---|---|---|---|---|---|---|
| $R\bar{3}m$-MoB$_3$ | 267 | 222 | 522 | 0.175 | 34.9 | 33.2 | - | 4.9 | 3.9 |
| $P6_3/mmc$-MoB$_3$ | 269 | 225 | 527 | 0.173 | 35.5 | 33.8 | - | 4.7 | 3.9 |
| $Pmmn$-MoB$_5$ | 264 | 232 | 538 | 0.16 | 38.6 | 37.3 | - | 5.6 | 3.8 |

**Higher molybdenum borides**

Our variable-composition evolutionary search reproduced most of the experimentally observed phases and predicted the boron-richest phase MoB$_5$ that has not been reported experimentally. There are several experimental reports of Mo–B phases with the composition close to MoB$_4$ and a controversial structure identification [39,40,46], whereas the evolutionary search did not find any structures with this composition that would match the experimentally observed XRD pattern.

Crystal structures in the boron-rich area of the Mo–B phase diagram are difficult to study experimentally because of a large difference in the X-ray scattering cross sections of boron and metal atoms which makes it hard to determine the exact positions of boron atoms and the boron content in the compound. Previous proposals for several structures of boron-rich Mo–B compounds have been made mainly on the basis of a chemical intuition and limited experimental data. These structures usually consist of sequences of hexagonal graphene-like boron and molybdenum layers with partial Mo occupancies and additional boron fragments of different forms. However, theoretical total energy computations showed that these hypothetical structures have high energies of formation, pointing at their instability. Galasso and Pinto [40] proposed the $P6_3/mmc$-MoB$_4$ phase with boron dimers placed between the hexagonal boron layers and oriented along the *c* axis of a unit cell. After the structural optimization of this phase, we obtained a positive energy of formation of 261 meV/atom, which agrees with previous theoretical works [11,12]. Another structure, suggested by Lundström and Rosenberg [39], has $Mo_{1-x}B_3$ composition with x ~ 0.20 and *P6₃/mmc* space group. It has similar crystal structure to *P6₃/mmc*-$MoB_3$, obtained during the USPEX evolutionary search, but with a portion of Mo atoms removed. To simulate this structure, we constructed a 2×2×1 supercell of *P6₃/mmc*-$MoB_3$ with three Mo vacancies in the Wyckoff position 2*b*, which resulted in the composition $Mo_{0.8125}B_3$. Its energy of formation is negative, -98 meV/atom, but lying quite high above the convex hull (Fig. 1), which makes this structure unstable. A recent experimental work considered a similar structure type with $Mo_{0.757}B_3$ composition [50]. We also examined the experiment-based structure of Nowotny et al. [41] for higher tungsten borides. This structure with the composition $Mo_{2-x}B_9$ features boron octahedra and lacks one of the two hexagonal graphene-like boron layers. After the structural optimization of $Mo_2B_9$, the atoms in the unit cell were significantly rearranged, but the resulting energy of formation remained positive at 114 meV/atom (Fig. 1). These examples show that the crystal structures proposed on the basis of the experimental measurements are unstable because of their high energies of formation. Although the positions of metal atoms can be robustly determined from XRD, constructing an energetically favorable arrangement of light boron atoms is extremely complicated. Experiment-independent structure prediction approaches have also been used before to resolve the correct crystal structure of boron-rich phases. The computations have revealed the *P6₃/mmc*-$MoB'_4$ phase [12] that also appeared in our global optimization procedure. This structure has a low energy of formation and is located only 15 meV/atom above the convex hull. However, its XRD pattern is distinctly different from that of the experimentally synthesized highest molybdenum boride [39,40,46]. In 2015, Lech et. al. carried out neutron diffraction studies of boron-rich W–B phases [6], partially replacing the tungsten atoms in the Wyckoff position 2*b* with boron triangles, which resulted in $WB_{4.2}$ chemical composition having P6₃/mmc space group. Similar structure patterns could be expected in the Mo–B phases because of their overall similarities with the W–B phases. However, our global optimization procedure did not reveal such structures with partial atomic occupations because of the limited unit cell size used in the evolutionary search (36 atoms per unit cell at the most). At the same time, it is possible to construct the energetically favorable Mo–B phases that have the compositions between $MoB_4$ and $MoB_5$ by analyzing the structures of USPEX-predicted phases.

The crystal structures of $P6_3/mmc$-$MoB_3$ and $Pmmn$-$MoB_5$ (Fig. 3a) have clear similarities: the alternation of graphene-like boron layers with the layers of Mo. The structure of $MoB_5$ can be produced from $MoB_3$ by replacing half of the Mo atoms in the Wyckoff position $2b$ by three boron atoms in a triangular arrangement (a $B_3$ unit), shown in orange in Fig. 3a. This agrees with the structure type proposed by Lech et al.[6]. We considered the possibility to construct stable $MoB_x$ phases having various compositions by replacing a different number of Mo atoms with boron triangles.

We first focused on the preferable position and orientation of an individual boron triangle. In the $P6_3/mmc$-$MoB_3$ structure, such $B_3$ units may occupy three Wyckoff positions: $2b$, $2c$, and $2d$ (Fig. 3b). To reveal the most favorable position and orientation of a $B_3$ unit, we constructed a 2×2×2 supercell of $P6_3/mmc$-$MoB_3$ and placed a single $B_3$ unit at these Wyckoff positions, either replacing the molybdenum atom at the corresponding site or filling an empty space of the Wyckoff position $2d$ (Fig. 3b). The two Mo layers in the $P6_3/mmc$-$MoB_3$ unit cell differ in rotation by $\frac{\pi}{3}$, therefore we considered the $B_3$ units in only one of these layers, keeping in mind that a change of layer would require a proper rotation of the boron triangle. We also considered a supercell with a single Mo vacancy in the *2b* position for comparison, since some of the experimentally suggested Mo-B phases have partial occupation of the Mo sublattice due to vacancies [39,50]. We relaxed the atomic coordinates and lattice parameters of the resulting supercells and evaluated their total energies using DFT. The distance above the convex hull $\Delta H$, or "metastability," was used as a quantitative measure to compare the relative stabilities of the structures. The resulting energies of formation as a function of composition are shown in Fig. 3c, the energy values are summarized in

Table 2. Most of the B$_3$ configurations, as well as a single Mo vacancy, have higher Δ*H* than the initial *P*6$_3$/*mmc*-MoB$_3$. The instability of the Mo vacancy is consistent with the high energy of formation of the Mo$_{0.8125}$B$_3$ structure, discussed above. The most stable configuration was realized when a boron triangle replaced a Mo atom in the Wyckoff position *2b* so as to maximize the distance to the closest in-plane Mo atoms (*2b*-I configuration). This most preferred B$_3$ configuration results in the lowering of Δ*H* compared to pure *P*6$_3$/*mmc*-MoB$_3$ from 18 to 14 meV/atom. Therefore, there is a prospect of the structure stabilization with the replacement of more Mo atoms in the Wyckoff positions *2b* and a consequent increase in the B$_3$ content. Such increase in the boron concentration would bring the energies of formation of the resulting structures closer to the convex hull line. However, as the number of B$_3$ units increases, the distance between them becomes smaller, their mutual position starts to play a role, and the interaction between the individual B$_3$ units has to be considered.

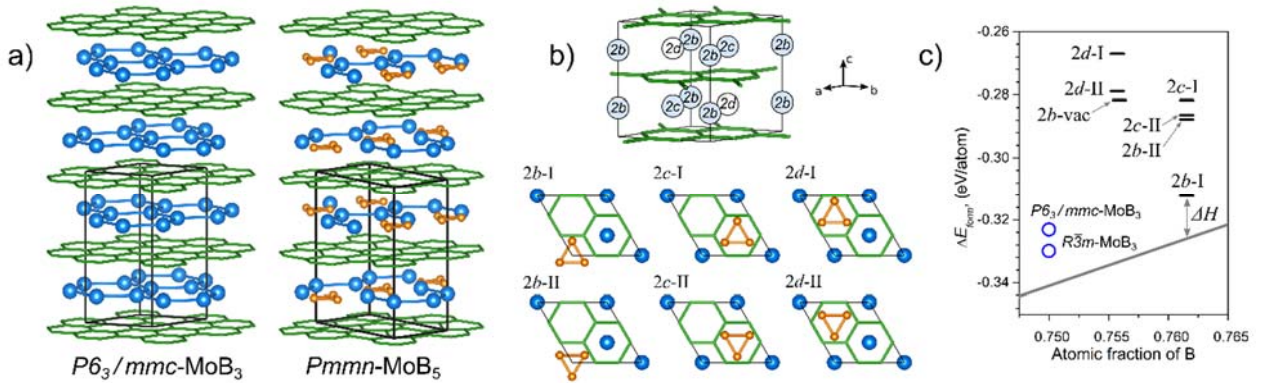

Fig. 3. (a) Crystal structures of *P*6$_3$/*mmc*-MoB$_3$ and *Pmmn*-MoB$_5$. To make the structure of each layer clear, the interlayer space is shown enlarged. The molybdenum atoms are shown as large blue circles. The hexagonal sheets of boron are depicted as a green wireframe. The boron atoms forming the triangular units are displayed in orange; (b) unit cell of *P*6$_3$/*mmc*-MoB$_3$ with the Wyckoff positions where a B$_3$ unit can be placed. Below, the top view of a single layer of MoB$_3$ with a B$_3$ unit placed in different positions and orientations; (c) energies of formation of 2×2×2 supercells with the B$_3$ unit in different configurations, together with a single Mo vacancy in the *2b* position (*2b*-vac). The convex hull is shown as a solid gray line connecting MoB$_2$ and MoB$_5$.

Table 2. Energies of formation and energy above the convex hull (Δ*H*) for 2×2×2 supercells of *P*6$_3$/*mmc*-MoB$_3$ with different locations of the B$_3$ unit and Mo vacancy in the *2b* position

| Structure type | $E_{form}$ (meV/atom) | Δ*H* (meV/atom) |
| --- | --- | --- |
| $P6_3/mmc$-MoB$_3$ | -323 | 18 |
| 2b-I | -312 | 14 |
| 2b-II | -288 | 38 |
| 2c-I | -282 | 44 |
| 2c-II | -286 | 40 |
| 2d-I | -267 | 67 |
| 2d-II | -279 | 55 |
| 2b-vac | -274 | 44 |
| $Pmmn$-MoB$_5$ | -233 | 0 |

The investigation of possible mutual arrangements of B$_3$ units requires consideration of many configurations with large supercells, complicating the computations that use the first-principles DFT approach. To approximate the DFT results, we developed a lattice model that is very accurate for the boron-rich Mo–B compounds. In this model, the Wyckoff positions 2b of $P6_3/mmc$-MoB$_3$ are considered as lattice sites that may be occupied by either Mo atoms or B$_3$ units in the 2b-I configuration. The total energy of a supercell consisting of $N$ unit cells of $P6_3/mmc$-MoB$_3$ with $0 <= n <= 2N$ B$_3$ units (there are 2$N$ possible sites for the placement of B$_3$ units because an individual $P6_3/mmc$-MoB$_3$ unit cell contains two Mo layers) can be approximately expressed as:

$$E_{N,n} = NE_{\text{MoB}_3} + n\Delta E_{\text{B}_3} + \frac{1}{2}\sum_{i \neq j} K_{ij}\, s_i s_j. \qquad (1)$$

In this formula, $E_{\text{MoB}_3}$ is the energy of a single $P6_3/mmc$-MoB$_3$ unit cell. $\Delta E_{\text{B}_3}$ is the energy change produced by the replacement of a single Mo atom by a B$_3$ unit in its lowest energy configuration, calculated as the total energy difference between a 2x2x2 supercell of $P6_3/mmc$-MoB$_3$ with a B$_3$ unit in the 2b-I configuration and a pristine 2x2x2 $P6_3/mmc$-MoB$_3$ supercell. The first two energy terms serve for the definition of the absolute energy value and may differ depending on the particular DFT implementation used. The third term represents the interaction between the individual B$_3$ units located in the 2b sites $i$ and $j$, where $K_{ij}$ is a parameter whose value depends on the distance between the sites that are occupied by the B$_3$ units. The interactions up to fourth nearest neighbor site were taken into account. Each 2b site has two nearest neighbors located above and below along the $c$ axis. Six second-order neighbors are located in the same layer plane. Twelve third-order neighbors occupy the upper and lower planes of 2b sites. The two remaining neighbors of the fourth order are located along the $c$ axis at a distance twice as long as the first-order neighbors. The graphical representation of the neighbor map is shown in Supporting Information Fig. S4. $s_i$ is equal to 1 or 0 if the $i$th site is occupied by a $B_3$ unit or a Mo atom, respectively. Thus, six parameters completely describe our lattice model, with two of them, obtained from the energies of 2x2x2 supercells, defining the absolute energy. Four remaining parameters, responsible for the $B_3$ interaction, were fit to reproduce the DFT total energies of supercells with multiple $B_3$ units. To do this, we constructed 18 2×2×2 and 14 2×2×3 supercells with randomly distributed $B_3$ units, having the compositions between $MoB_{3.4}$ and $MoB_{5.23}$. We calculated their total energies within the DFT approach and then minimize the squared difference between the DFT energies and lattice model energies from eq. (1) with use of the random walk minimization procedure. The resulting set of parameters is presented in Table 3. All parameters responsible for the interaction have positive values, which indicates the repulsive nature of the $B_3$–$B_3$ interaction. For the supercells constructed for the fitting procedure, the largest absolute difference between the total energy calculated using the DFT and our lattice model was ~1 meV/atom. By construction, the highest accuracy of our lattice model is attained for boron-poor systems, diminishing as the boron content increases (the largest error in the total energy was 12 meV/atom for $MoB_9$ structure).

Table 3. Interaction parameters used for calculating the total energies within the lattice model (1).

| Neighbor order | Number of neighbors | Distance | $K$ (eV) |
|---|---|---|---|
| 1 | 2 | $c/2$ | 1.289 |
| 2 | 6 | $a$ | 0.095 |
| 3 | 12 | $\sqrt{a^2 + c^2/4}$ | 0.019 |
| 4 | 2 | $c$ | 0.103 |

Having accurately parameterized the lattice model, we searched for the most stable composition, corresponding to the optimal arrangement of the $B_3$ units, by considering the 2×2×3 $P6_3/mmc$-$MoB_3$ supercell and performing a brute-force search for lowest-energy structures, taking into account all possible arrangements of the $B_3$ units. The energies of formation of the constructed nonstoichiometric structures are shown in Fig. 4a, displaying only the boron-rich region of the convex hull diagram from Fig. 1. An increase in the boron content leads to a decrease in the energy of formation, which approaches the convex hull line at the composition close to MoB$_4$ (within the 2x2x3 supercell approach the obtained composition is MoB$_{4.02}$, lying 2 meV/atom above the convex hull) and closely follows this line until the composition MoB$_5$ on the convex hull is reached. Further increase in the boron content leads to a rapid decrease in stability (the energy of formation increases). The composition range between $4 \lesssim x \leq 5$ is therefore the most stable within the considered structure type, with MoB$_5$ being the most boron rich among them. To verify the accuracy of the obtained results, we recalculated the energy of formation of some structures within the DFT method. The corresponding energies of formation, marked by squares in Fig. 4a, show an excellent agreement between the lattice model and DFT calculations (a direct comparison between the results of the lattice model and DFT is shown in Fig. 4b). Fig. 4c shows the dependence of the equilibrium lattice constants (per single *P*6$_3$/*mmc* primitive cell) on the boron content of the lowest-energy MoB$_x$ phases. Such dependence could be used for experimental determination of the chemical composition of these compounds, by experimentally calibrating the dependence of the cell parameters on the composition. The plot clearly shows different behavior at different boron concentrations. In the range from MoB$_3$ to MoB$_{4.2}$, the addition of more B$_3$ units leads to a small increase of both lattice constants *a* and *c*. For structures between MoB$_{4.2}$ to MoB$_5$, which correspond to the interval of the most stable compositions, the parameter *c* increases from 6.31 Å to 6.37 Å with a small decrease of lattice constant *a* from 5.21 Å to 5.20 Å. Further increase of the boron content results in an opposite trend, with a fast decrease of the *c* parameter and increase of *a*.

During the brute-force search, we collected the data on the highest-energy structures to define the origins of the energy disadvantage in terms of the B$_3$ units arrangement. The crystal structures of the least and most stable phases are shown in Fig. 4d and e to clearly distinguish the structural difference between them. In the low-energy structures (Fig. 4e), the B$_3$ units, shown in orange, are uniformly distributed within the considered cells, whereas in the high-energy structures (Fig. 4d), the B$_3$ units tend to form vertical columns. High energy of these structures is the consequence of the repulsion between the B$_3$ units, which also explains the jump in the energy for structures whose atomic fraction of boron is higher than 5/6. MoB$_5$ is the boron-richest phase where the B$_3$ units can be distributed without being placed in the adjacent sites. It is the highest theoretically achievable molybdenum boride within the proposed type. The lowest-energy B$_3$ arrangements, presented in Fig. 4e, are not unique. Multiple different configurations, corresponding to the ground state energies, were found during our brute-force search, which points at the disordered nature of this material. However, at the MoB$_5$ composition, the strong repulsive interaction between the nearest neighbor B$_3$ units reduces the number of lowest energy configurations and suggests the ordering of B$_3$ units at compositions close to MoB$_5$.

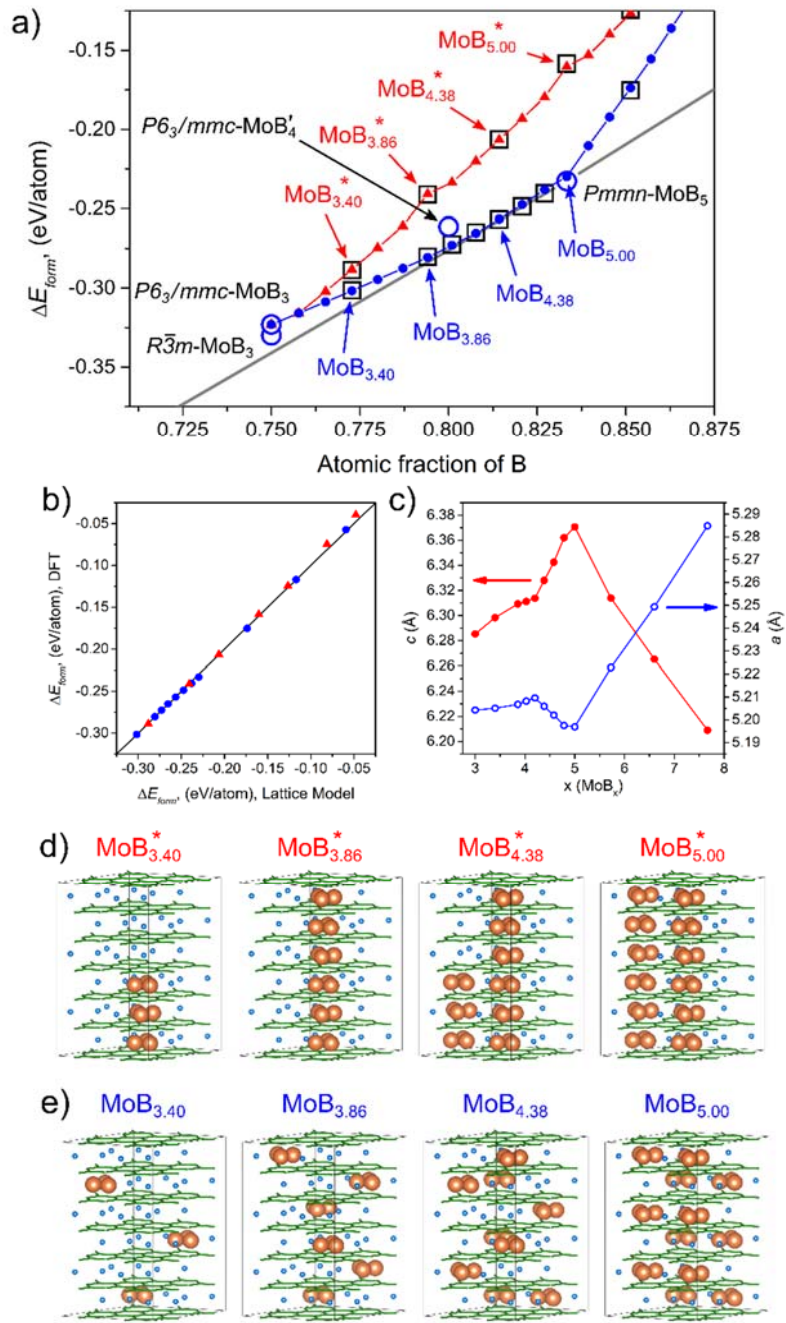

Fig. 4. (a) Energies of formation of boron-rich MoB$_{3-9}$ structures. A gray line depicts the convex hull. Structures obtained during the evolutionary global search are shown as hollow circles. The lowest- and highest-energy structures obtained using the lattice model are shown as small blue circles and red triangles, respectively. Hollow squares show the energies of structures recalculated within the DFT. (b) comparison between the energies of formation for the most and least stable MoB$_x$ phases, obtained within the lattice model and recalculated within DFT. (c) Dependence of the equilibrium lattice constants per single $P6_3/mmc$ primitive cell on the composition of the low-energy MoB$_x$ phases. (d) Selected structures with the highest energy of formation showing the energetically unfavorable columns of the B$_3$ units. The boron atoms forming the triangular units are displayed in orange. (e) Lowest-energy structures of the same compositions, the B$_3$ units are uniformly distributed in the bulk of a crystal.

Our calculations show that the MoB$_x$ structures, constructed via partial replacement of the *2b*-Mo atoms of the *P*6$_3$/*mmc*-MoB$_3$ phase by triangular B$_3$ units, are energetically more preferable than the previously proposed boron-rich Mo-B phases with compositions close to MoB$_4$, including *P*6$_3$/*mmc*-MoB$_4'$ (see Fig. 4a) and *P*6$_3$/*mmc*-Mo$_{1-x}$B$_3$ (see Fig. 1 and Fig. 3c and Ref. [39]). We compared the simulated X-ray diffraction patterns for these molybdenum borides (Fig. 5). The XRD pattern of *P*6$_3$/*mmc*-MoB$_4'$ (Fig. 5a) is clearly different from the most widely accepted *P*6$_3$/*mmc*-Mo$_{1-x}$B$_3$ (x=0.20) model (Fig. 5b), proposed experimentally [39]. To obtain the XRD pattern of the MoB$_x$ phases within our suggested structure type, we considered a 12x12x10 supercell within our lattice model (1) and performed an annealing of the B$_3$ positions at 2000 K with the Metropolis Monte Carlo algorithm to simulate the high-temperature synthesis conditions. The resulting XRD pattern for the MoB$_4$ composition (Fig. 5c) is in good agreement with the experimental model. The main difference between them is related to the peak at 19.7°, whose intensity is determined the value of the fractional occupation of the Mo atoms in the *2b* position. The replacement of a larger fraction of Mo atoms with B$_3$ units, which leads to the composition of MoB$_{4.7}$ (Fig. 5d), allowed us to decrease the relative intensity of this peak, bringing the resulting XRD spectrum in closer agreement with those of the experimental Mo$_{0.8}$B$_3$ structure. Further increase of boron content resulted in the appearance of additional peaks around 18° in the simulated XRD pattern because of the B$_3$ units ordering. Such ordering appears due to the strong repulsive interaction between the B$_3$ units, located at the adjacent sites, which restricts the number of their possible arrangements.

Good agreement between the experimental XRD pattern [39] and the XRD pattern of our proposed structure, together with its favorable thermodynamics, gives strong evidence in favour of this structure (which can be described as based on the *P*6$_3$/*mmc* -MoB3 structure with partial replacement of Mo atoms in the *2b*-position with triangular B$_3$ units). Our calculations show that the highest theoretically achievable molybdenum boride within this structure type is MoB$_5$.

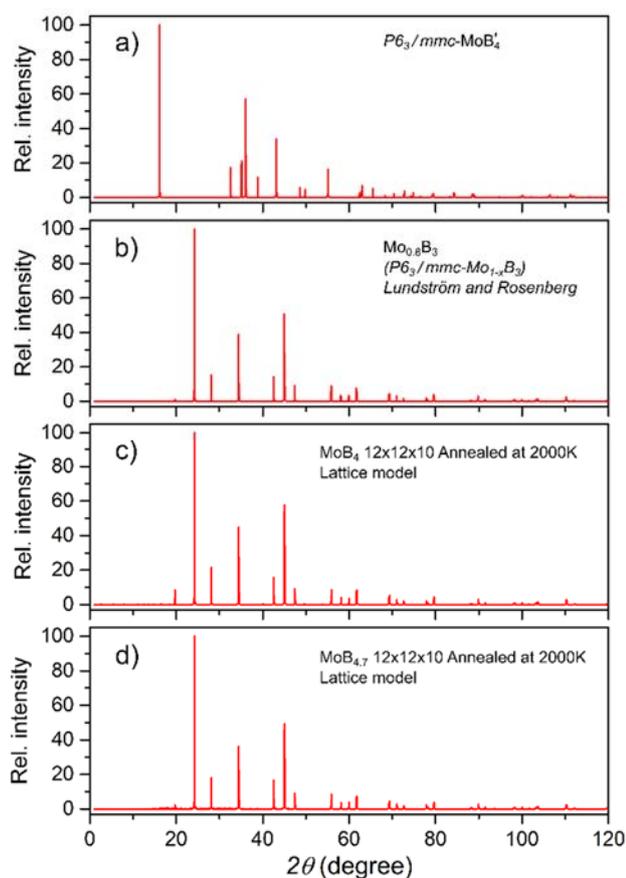

Fig. 5. Simulated XRD spectra. a) Previously theoretically proposed $P6_3/mmc$-MoB$_4'$. b) Experimentally proposed $P6_3/mmc$-Mo$_{1-x}$B$_3$ model [39] with $x=0.2$. c) MoB$_4$ and d) MoB$_{4.7}$ structures, constructed as 12x12x10 supercells of the $P6_3/mmc$-MoB$_3$ phase, with a partial replacement of the Mo atoms in the *2b*-position by B$_3$ units.

In conclusion, the crystal structure of boron-rich Mo–B phases has been debated for a long time. The structural complexity of pure boron manifests itself in the boron-rich part of the transition metal boride phase diagram. In this work we used a range of computational approaches, including the evolutionary global optimization, first-principles total energy calculations, and a parameterized lattice model to investigate the atomic structures of Mo–B phases in the whole compositional range. Beside the phases reported and characterized in previous experimental and theoretical studies, we found a new stable phase of MoB$_5$, having the same structure type as previously predicted WB$_5$ [8]. We showed that the appropriate mixing of structural fragments of MoB$_5$ and metastable $P6_3/mmc$-MoB$_3$ (the partial replacement of the molybdenum atoms in the Wyckoff position *2b* by appropriately rotated boron triangles B$_3$) makes it possible to construct a broad range of structures with the compositions between MoB$_3$ and MoB$_9$. The development and application of a lattice model for the total energy calculations allowed us to find energetically stable boron-rich phases with MoB$_4$–MoB$_5$ compositions. The energetically favorable MoB$_x$ phases correspond to a uniform distribution of the B$_3$ units within the hexagonal lattice, based on the tendency of these units to avoid each other in the bulk. The crystal structure of MoB$_x$ with low concentrations of B$_3$ units is disorder and similar to those proposed by Lech et al. [6] for the W–B system. An increase in the boron content decreases spatial freedom for boron triangles due to the repulsive interaction neighboring B$_3$ units, resulting in their ordering at compositions close to MoB$_5$. The estimations of Vickers hardness show that higher borides have a high hardness of 33–39 GPa, indicating possible superhardness of these materials. In addition, we showed that the computational structure prediction methods are powerful tools for determining the structure of systems in situations where an experimental crystal structure assessment is complicated.

## Acknowledgments
This work was supported by the Russian Science Foundation (№ 19-72-30043). The calculations were carried out on Rurik supercomputer at MIPT and Arkuda and Pardus supercomputers of the Skolkovo Foundation.

# Supporting Information

# Structure, Stability and Mechanical Properties of Boron-Rich Mo–B Phases: A Computational Study

*Dmitry V. Rybkovskiy,[1,2] Alexander G. Kvashnin,[1,3,*] Yulia A. Kvashnina,[4] and Artem R. Oganov[1,3,5]*

[1] Skolkovo Institute of Science and Technology, Skolkovo Innovation Center, 3 Nobel Street, Moscow 121205, Russia
[2] A. M. Prokhorov General Physics Institute of RAS, 38 Vavilov Street, Moscow 119991, Russia
[3] Moscow Institute of Physics and Technology, 9 Institutsky Lane, Dolgoprudny 141700, Russia
[4] Pirogov Russian National Research Medical University, 1 Ostrovityanova Street, Moscow 117997, Russia
[5] International Center for Materials Discovery, Northwestern Polytechnical University, Xi'an 710072, China

**Corresponding Author**
*A.G. Kvashnin, E-mail: A.Kvashnin@skoltech.ru

Content

# Crystal data

Table S1. Crystal structures of predicted Mo-B phases.

| Phase | Volume Å³/atom | Lattice parameters | Coordinates | | | |
|---|---|---|---|---|---|---|
| | | | atom | x | y | z |
| $I4/mcm$-Mo$_2$B | 12.23 | a = 5.55 Å<br>c = 4.77 Å<br>a = 5.54 Å [1]<br>c = 4.74 Å [1]<br>a = 5.54 Å [2]<br>c = 4.73 Å [2] | Mo1<br>B1 | 0.170<br>0.000 | 0.330<br>0.000 | 0.000<br>0.250 |
| $I4/m$-Mo$_2$B | 12.24 | a = 5.56 Å<br>c = 4.75 Å | Mo1<br>B1 | 0.177<br>0.000 | 0.337<br>0.000 | 0.000<br>0.212 |
| $P4/mbm$-Mo$_3$B$_2$ | 11.45 | a = 6.03 Å<br>c = 3.15 Å | Mo1<br>Mo2<br>B1 | -0.328<br>0.000<br>0.113 | 0.172<br>0.000<br>0.613 | 0.500<br>0.000<br>0.000 |
| $I4_1/amd$-MoB<br>(α-MoB) | 10.38 | a = 3.12 Å<br>c = 17.01 Å<br>a = 3.11 Å [1]<br>c = 16.97 Å [1]<br>a = 3.10 Å [2]<br>c = 16.95 Å [2] | Mo1<br>B1 | 0.000<br>0.000 | 0.250<br>0.250 | 0.072<br>0.220 |
| $Cmcm$-MoB<br>(β-MoB) | 10.39 | a = 3.16 Å<br>b = 8.53 Å<br>c = 3.09 Å<br>a = 3.16 Å [1]<br>b = 8.61 Å [1]<br>c = 3.08 Å [1] | Mo1<br>B1 | 0.000<br>0.000 | 0.144<br>0.440 | 0.250<br>0.250 |
| $R\bar{3}m$-MoB$_2$ | 9.18 | a = 3.02 Å<br>c = 20.94 Å<br>a = 3.019 Å [3]<br>c = 20.961 Å [3] | Mo1<br>B1<br>B2 | 0.000<br>0.000<br>0.000 | 0.000<br>0.000<br>0.000 | 0.076<br>0.182<br>0.332 |
| $P6_3/mmc$-MoB$_2$ | 9.18 | a = 3.02 Å<br>c = 13.96 Å | Mo1<br>B1<br>B2<br>B3 | 0.333<br>0.333<br>0.333<br>0.000 | 0.667<br>0.667<br>0.667<br>0.000 | -0.364<br>0.023<br>0.250<br>0.250 |
| $P6/mmm$-MoB$_2$ | 8.80 | a = 3.03 Å<br>c = 3.33 Å | Mo1<br>B1 | 0.000<br>0.333 | 0.000<br>0.667 | 0.000<br>0.500 |
| $R\bar{3}m$-Mo$_2$B$_5$ | 8.38 | a = 3.08 Å<br>c = 21.39 Å<br>a = 3.01 Å [1]<br>c = 20.93 Å [1]<br>a = 3.1 Å [2]<br>c = 20.92 Å [2] | Mo1<br>B1<br>B2<br>B3 | 0.000<br>0.000<br>0.000<br>0.000 | 0.000<br>0.000<br>0.000<br>0.000 | 0.076<br>0.331<br>0.185<br>0.500 |
| $P6_3/mmc$-Mo$_2$B$_5$ | 8.80 | a = 3.01 Å<br>c = 15.66 Å | Mo1<br>B1<br>B2<br>B3 | 0.000<br>0.333<br>0.333<br>0.000 | 0.000<br>0.667<br>0.667<br>0.000 | -0.102<br>-0.198<br>-0.497<br>0.250 |
| $R\bar{3}m$-MoB$_3$ | 9.20 | a = 5.22 Å<br>c = 9.36 Å<br>a = 5.224 Å [4]<br>c = 9.363 Å [4] | Mo1<br>B1 | 0.000<br>-0.335 | 0.000<br>0.000 | -0.165<br>0.000 |
| $P6_3/mmc$-MoB$_3$ | 9.22 | a = 5.20 Å<br>c = 6.29 Å<br>a = 5.208 Å [3]<br>c = 6.290 Å [3] | Mo1<br>Mo2<br>B1 | 0.000<br>0.333<br>-0.335 | 0.000<br>0.667<br>0.000 | 0.250<br>0.750<br>0.000 |
| $P6_3/mmc$-MoB$'_4$ | 8.27 | a = 2.95 Å<br>c = 10.98 Å<br>a = 2.951 Å [4]<br>c = 10.983 Å [4] | Mo1<br>B1<br>B2 | 0.333<br>0.333<br>0.333 | 0.667<br>0.667<br>0.667 | 0.750<br>-0.455<br>0.111 |
| | | | Mo1 | 0.250 | 0.250 | -0.418 |
| | | | Mo2 | 0.250 | 0.750 | 0.252 |
| | | | Mo3 | 0.250 | 0.750 | -0.089 |
| *Pmmn*-MoB$_5$ | 8.27 | a = 5.19 Å<br>b = 6.37 Å<br>c = 9.01 Å | B1 | 0.081 | 0.493 | 0.416 |
| | | | B2 | -0.417 | 0.494 | 0.252 |
| | | | B3 | 0.415 | 0.494 | 0.082 |
| | | | B4 | 0.431 | 0.250 | 0.307 |
| | | | B5 | 0.250 | 0.250 | 0.129 |

## Mechanical properties

The Ashby plot of Vickers hardness vs fracture toughness was constructed (see Fig. S1), which allows one to clearly find materials with an optimal combination of Vickers hardness and fracture toughness. Phases predicted here are denoted by red circles, while black and blue points correspond to known materials (diamond, α-B, c-BN, WC, TiN, CrB$_4$ etc.) and W-B phases from Ref. [5], respectively. The best combination of hardness and fracture toughness, if one excludes high pressure phases (diamond and cubic BN), belongs to CrB$_4$, WB$_5$ and WC. ZrB$_2$, WB$_2$ together with W$_4$B$_7$ lie on the second Pareto front. MoB$_3$ and MoB$_5$, predicted here, and $P\bar{4}2_1/m$-WB, are third best choice. MoB$_5$ seems superior to the widely used TiN, as well as to pure B and B$_4$C. MoB$_5$ displays highest hardness among studied Mo-B phases with the hardness of 38 GPa with comparably high fracture toughness.

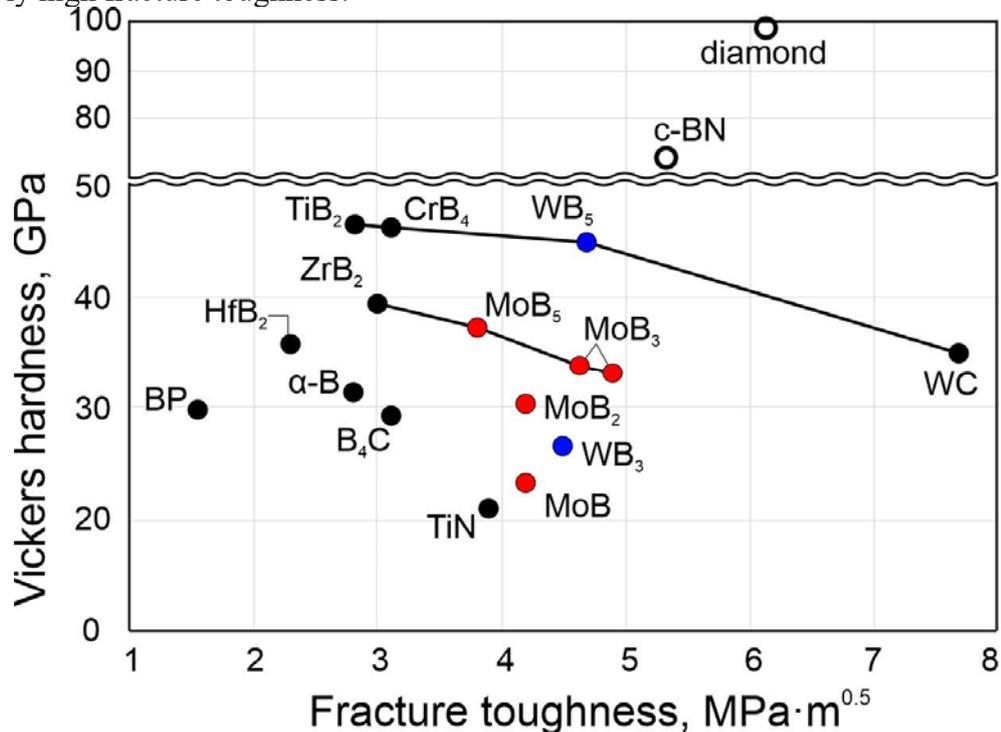

Fig. S1. Ashby plot of Vickers hardness vs. fracture toughness for predicted Mo-B phases (red points) compared with known materials (black points) and W-B phases from Ref. [5] (blue points). Horizontal line corresponds to Vickers hardness of WC. Black lines correspond to third and fourth Pareto fronts, while diamond and c-BN are first and second Pareto fronts, respectively. Vickers hardness and fracture toughness were calculated using Mazhnik-Oganov model [6].

**Physical properties of MoB$_5$**

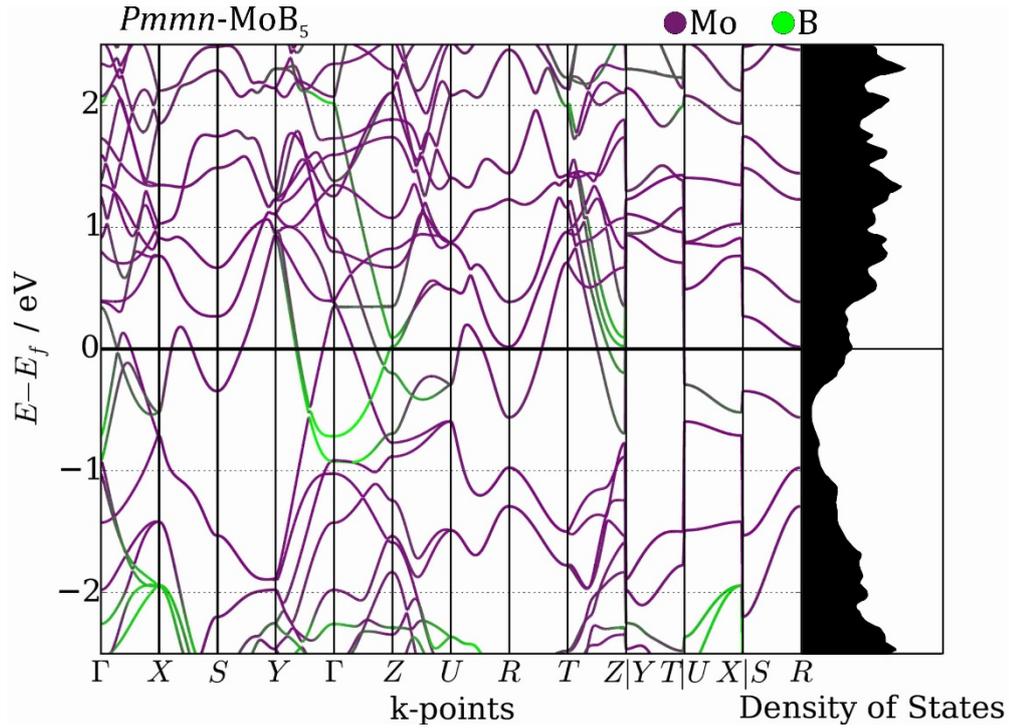

Fig. S2. Atom-projected bands structure of MoB$_5$ together with DOS. Violet color is for Mo, green – boron.

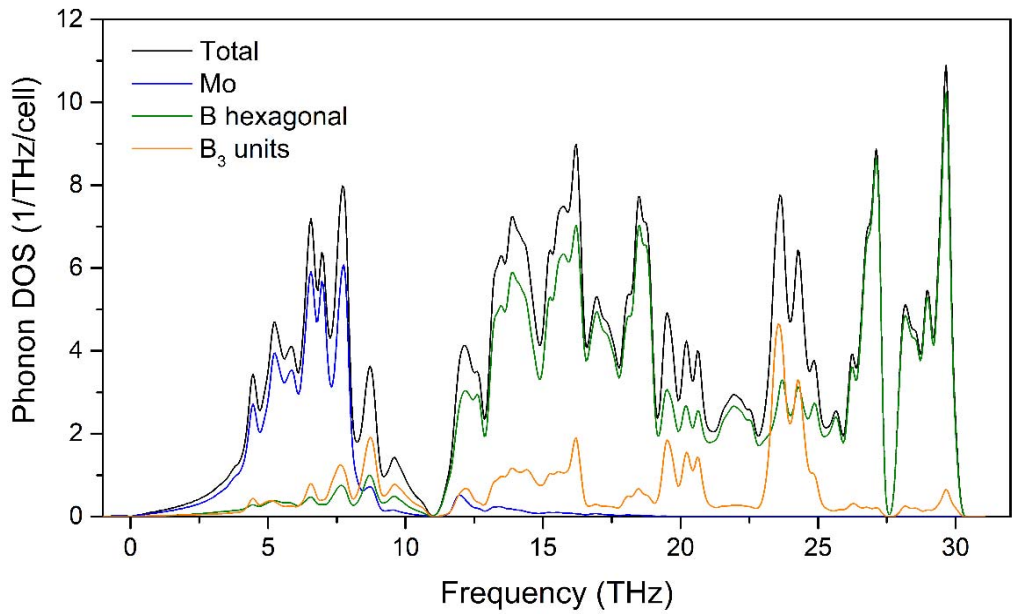

Fig. S3. Phonon density of states for *Pmmn*-MoB$_5$ unit cell with 36 atoms and volume of 297.87 Å$^3$. Color lines show contributions from molybdenum atoms, hexagonal boron wireframe and triangular boron units.

## Lattice model neighbor map

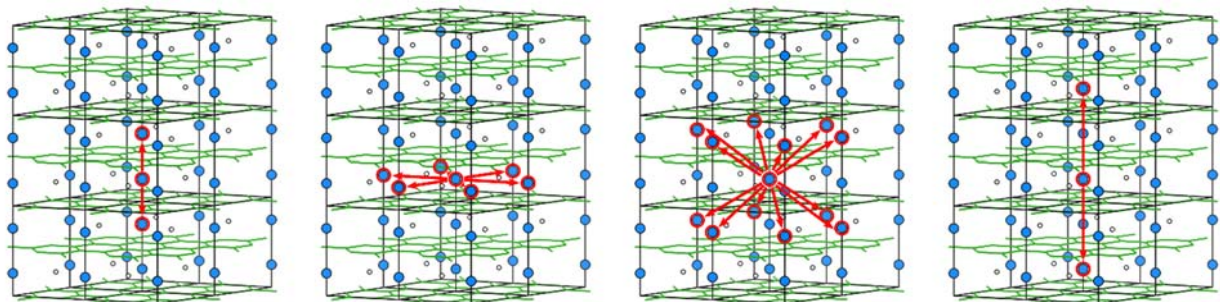

Fig. S4. Nearest neighbors used to consider the interaction between neighboring $B_3$ units.